\documentclass[runningheads]{llncs}
\usepackage[T1]{fontenc}
\usepackage{graphicx}
\usepackage{subcaption} 
\usepackage{booktabs}
\usepackage{tabularx}
\usepackage{multirow}   
\usepackage{makecell}   
\usepackage[dvipsnames]{xcolor}
\definecolor{ForestGreen}{rgb}{0.13, 0.55, 0.13}

\usepackage{import} 
\usepackage{threeparttable}
\usepackage{amsmath}
\usepackage{hyperref}
\usepackage{soul,color}
\soulregister\cite7
\soulregister\ref7
\soulregister\pageref7
\setstcolor{red}
\setul{}{.2ex}
\hypersetup{
    colorlinks=true,
    linkcolor=blue,
    citecolor=blue,
    urlcolor=blue
}
\usepackage{bbm}
\usepackage{enumitem}
\usepackage[numbers, sort]{natbib}
\usepackage[expansion,protrusion]{microtype}

\begin{document}

%
\title{Query–Document Dense Vectors for \\ LLM Relevance Judgment Bias Analysis}
%
%
\vspace{-25mm}
\author{Samaneh Mohtadi\inst{1}\orcidID{0009-0003-0980-6254} \and
Gianluca Demartini\inst{1}\orcidID{0000-0002-7311-3693}
}
\authorrunning{S. Mohtadi and G. Demartini}
\institute{The University of Queensland, Brisbane, Australia\\
\email{s.mohtadi@uq.edu.au, g.demartini@uq.edu.au}\\}
\maketitle              
\begin{abstract}
Large Language Models (LLMs) have been used as relevance assessors for Information Retrieval (IR) evaluation collection creation due to reduced cost and increased scalability as compared to human assessors.
While previous research has looked at the reliability of LLMs as compared to human assessors, in this work, we aim to understand if LLMs make systematic mistakes when judging relevance, rather than just understanding how good they are on average. To this aim, we propose a novel representational method for queries and documents that allows us to analyze relevance label distributions and compare LLM and human labels to identify patterns of disagreement and localize systematic areas of disagreement. We introduce a clustering-based framework that embeds query–document (Q–D) pairs into a joint semantic space, treating relevance as a relational property. Experiments on TREC Deep Learning 2019 and 2020 show that systematic disagreement between humans and LLMs is concentrated in specific semantic clusters rather than distributed randomly. Query-level analyses reveal recurring failures, most often in definition-seeking, policy-related, or ambiguous contexts. Queries with large variation in agreement across their clusters emerge as disagreement hotspots, where LLMs tend to under-recall relevant content or over-include irrelevant material. This framework links global diagnostics with localized clustering to uncover hidden weaknesses in LLM judgments, enabling bias-aware and more reliable IR evaluation.

\keywords{Relevance Judgments \and Bias Analysis \and Information Retrieval Evaluation \and Semantic Clustering \and Large Language Models. }
\end{abstract}
\section{Introduction}
Large Language Models (LLMs) are increasingly being used as automatic judges for Information Retrieval (IR) evaluation. Recent studies show that while LLMs often align with human relevance labels, certain disagreement remains. 
The evaluation of IR systems has long relied on relevance judgments, decisions about whether a document is useful for answering a given query. These judgments are fundamental, as they determine how search systems are ranked and assessed~\cite{lawrence1998pagerank,pan2007google}. As an example, a standard TREC track requires six expert assessors dedicated full-time over several weeks ~\cite{soboroff2025don}. Beyond expense, human judgments are also known to exhibit variability, being order dependent, and inconsistent across assessors~\cite{alaofi2024llms}. To address these challenges, researchers have begun exploring whether LLMs, which produce fluent text in response to natural-language prompts, could serve as automated assessors~\cite{bauer2023sigirforum,faggioli2024determines}. LLMs have already been applied to predict relevance in TREC collections and commercial search results demonstrating much higher throughput and lower annotation cost than human judges (including crowd workers)~\cite{thomas2024large}. Importantly, several studies report that LLM-based judgments, while not flawless, achieve a level of alignment with human assessments that is often sufficient for system-level evaluation~\cite{balog2025rankers,faggioli2023perspectives,faggioli2024determines,thomas2024large,upadhyay2024large}. LLMs also appear to be less affected by context switching, often producing more consistent decisions across changing evaluation conditions~\cite{frobe2025large}. Moreover, LLMs have demonstrated substantial promise in automating the evaluation of IR systems, providing relevance labels that closely approximate human annotation quality~\cite{li2025generation,thomas2024large,zhuang2024beyond}.

Despite these advantages, LLM-based judgments may still suffer from biases, reliability vulnerabilities, and misalignment with subtle human judgments~\cite{li2025generation,zhuang2024beyond}. For example, Alaofi et al.~\cite{alaofi2024llms}  found that LLMs are particularly prone to false positives when query terms appear in candidate passages, suggesting an over-reliance on surface lexical overlap. Similarly, Zhuang et al.~\cite{zhuang2024beyond} caution that binary “Yes/No” prompts can produce noisy or biased outputs for documents that are only partially relevant, since such cases lack intermediate label options. Rahmani et al.~\cite{rahmani2025towards} further show that synthetic test collections constructed with LLMs can introduce systematic biases, which then propagate into downstream evaluation results. As Faggioli et al.~\cite{faggioli2023perspectives} emphasize, evaluations in IR must ultimately remain grounded in human judgment to preserve trust and reliability, yet ongoing debates continue over how closely LLM judgments align with expert annotators~\cite{balog2025rankers,dietz2025llm}. While these studies provide valuable insights into the reliability of LLMs, conventional research has largely focused on detecting whether such biases exist, rather than localizing \textit{where} they emerge, leaving open the question of how to map disagreement to specific semantic contexts.

This paper aims to diagnose hidden biases in LLM relevance judgments, improve transparency, and strengthen reliability. We embed query–document (Q–D) pairs as dense vectors into a joint semantic space, treating relevance as a relational property rather than as separate query and document embeddings. By clustering this embedding space, semantically similar Q–D pairs are grouped into coherent neighborhoods for analysis. This clustering-based agreement analysis provides richer insights than global agreement statistics, enabling us to detect \textit{where} human and LLM annotators systematically disagree. Specifically, we pursue the following research questions:
\begin{enumerate}[label={RQ\arabic*},leftmargin=*,noitemsep,topsep=0pt]
    \item \label{rq:embedding} To what extent do joint query–document embeddings capture human-LLM relevance interactions?
    
    \item \label{rq:bias} Where and why do LLM relevance judgments systematically diverge from human judgments across semantic clusters?
\end{enumerate}
Our contributions are as follows:
\begin{enumerate}[label={C\arabic*},leftmargin=*]
    \item \label{contrib:c1} We propose a clustering-based framework for analyzing agreement and bias in LLM relevance judgments by embedding Q–D pairs into a joint semantic space, moving beyond global statistics to local, context-aware diagnostics.
    
    \item \label{contrib:c2} We use Gwet’s AC1~\cite{gwet2008computing} for IR evaluation as a chance-adjusted agreement measure. To our knowledge, this is the first use of AC1 in IR, addressing the limitations of widely used measures such as Cohen’s $\kappa$ by providing more stable estimates under label imbalance and enabling finer-grained cluster-level agreement analysis.
    
   \item \label{contrib:c3} We propose cluster-based agreement variation as a measure of how LLM– human agreement for the same query varies across different semantic neighborhoods. This measure allows us to localize systematic disagreement, distinguish stable from unstable regions of the embedding space, and identify queries that are especially bias-prone, providing diagnostics of semantic blind spots.

\end{enumerate}
\section{Related Work}
The wider research landscape has increasingly examined the validity and limitations of LLM-as-judge pipelines. Benchmarking studies showed that LLMs can approximate user preferences when carefully prompted~\cite{thomas2024large}, providing initial evidence of their potential for IR evaluation. Other work, however, raises concerns about over-reliance, noting risks of bias, limited transparency, and the need for rigorous methodology~\cite{clarke2024llm,dietz2025llm,soboroff2024dont}. At the same time, surveys and position papers warn about the risks of bias reinforcement, reproducibility challenges, and methodological inconsistency when relying on LLMs as evaluators~\cite{dietz2025llm,faggioli2023perspectives}.
Empirical error analyses have further exposed weaknesses in LLM judgments. Alaofi et al.~\cite{alaofi2024llms} show that models can be misled by lexical overlap and adversarial cues, and over-inference, and observe that LLMs tend to over-predict relevance while their non-relevance labels are more conservative and dependable. Comparative studies find that LLM assessors often agree more with each other than with human judges~\cite{frobe2025large}. In the TREC 2024 RAG track, Upadhyay et al.~\cite{upadhyay2024umbrela} argued that automated judgments could replace human labels, but this claim was soon contested. Clarke and Dietz~\cite{dietz2025llm} highlighted risks of circularity and manipulation, while Balog et al.~\cite{balog2025rankers} provided empirical evidence that LLM assessors exhibit bias, tending to prefer runs produced by other LLMs.

These findings highlight the need to understand not only \textit{how much} disagreement occurs, but also \textit{where and why} it arises in the semantic space of queries and documents. Prior benchmark efforts~\cite{arabzadeh2025benchmarking,dietz2025llm,li2024llms} focus on macro-level statistics that average over diverse judgments, leaving open how disagreement is structured and where it originates. 

More recently, research has shifted toward diagnosing bias and validity in synthetic evaluation.  Rahmani et al.~\cite{rahmani2025towards} show that synthetic queries diverge from human ones in style and length, while GPT-4 judgments are systematically lenient, introducing a positive bias that inflates absolute performance scores while leaving relative rankings comparatively stable. Arbabi et al.~\cite{arbabi2025relative} introduce the Relative Bias framework, combining embedding analysis with LLM-as-judge evaluation to quantify over- and under-labeling tendencies across topical subspaces. Together, these studies show that bias is structured and context-dependent rather than random. These findings underscore the importance of examining not just \textit{how much} disagreement occurs, but also \textit{where} and \textit{why} it arises, an open question that global IR evaluation metrics cannot adequately address.

In IR evaluation, agreement reliability is crucial since system effectiveness depends on the reliability of relevance judgments. Cohen’s $\kappa$~\cite{cohen1960coefficient} has been the dominant measure~\cite{arabzadeh2025benchmarking,frobe2025large,cohen1960coefficient}, but it is unstable under class imbalance, a problem referred to as the “$\kappa$ paradox”~\cite{cicchetti1990kappa}. In imbalanced datasets, $\kappa$ may underestimate agreement even when assessors substantially agree, because much of that agreement is discounted as chance. Beyond IR, Vidgen et al.~\cite{vidgen2021hatespeech} applied Gwet’s AC1 in hate speech annotation, showing its robustness under skewed labels. Haley et al.~\cite{haley2008using} highlighted AC1’s advantages as a suitable alternative for IR contexts, though it has not been widely adopted. This indicates awareness of AC1’s advantages, but its systematic use in IR evaluation remains unexplored.

Classical IR research established that relevance is inherently subjective, shaped by intent, task, and ambiguity~\cite{cuadra1967opening,rees1967field}. This subjectivity motivates treating relevance not as a query-only or document-only property, but as a relational property of the query–document pair. Recent work highlights the need for semantically grounded methods that can reveal not only how much misalignment exists between LLMs and humans, but also where and why it occurs. 
++
Our study advances this line of inquiry by moving beyond label-level comparisons toward an embedding-based approach that localizes disagreement within a semantic space. By analyzing neighborhoods in the joint Q–D embedding space, examining cluster-level agreement stability, and identifying bias patterns, we expose latent structures of misalignment that corpus-level averages conceal.
\section{Methodology and Experimental Setting}
We outline our methodology and setting for analyzing relevance judgments through semantic clustering of query-document (Q–D) pairs.~\footnote{For replicability purposes, all embedding model settings and instructions are available online at: [REDACTED FOR BLIND REVIEW].}
\subsection{Datasets}
We evaluate on benchmark datasets of TREC Deep Learning 2019 (DL-2019)~\cite{craswell2020overviewtrec2019deep} and TREC Deep Learning 2020 (DL-2020)~\cite{craswell2021overviewtrec2020deep}, built on the MS MARCO passage collection~\cite{bajaj2016ms}, together with LLM relevance judgments released by Fr{\"o}be et al.~\cite{frobe2025large}. As representative LLM judges, we select Claude-3-haiku, Gemini-1.5-flash-8b, GPT-4o, and Llama-3.1, since Fr{\"o}be et al. report that these models consistently achieve strong agreement, similarity, and ranking performance. Table~\ref{tab:trec_datasets} summarizes the characteristics of the datasets. Our experiments are based on a binary relevance setting, and thus, graded qrels are binarized by mapping scores of 2 and 3 to relevant (1) and scores of 0 and 1 to non-relevant (0), aligned with TREC’s recommendation and the protocol~\cite{damessie2017gauging}.
\begin{table}[h]
    \centering
    \scriptsize
    \caption{TREC Dataset Characteristics}
    \label{tab:trec_datasets}
    \begin{tabular}{cccccc}
    \toprule
    Dataset Name & Queries	& Documents & Relevance Judgments &  Relevant(\%) &  Not-Relevant(\%) \\
    \midrule
    DL-2019& 43 & 9139 & 9260 & 27.0 & 73.0 \\
    DL-2020 & 54 & 11224 & 11386 & 15.0 & 85.0 \\
    \bottomrule
\end{tabular}    
\end{table}
\subsection{Embedding Q–D Pairs}
We adopt INSTRUCTOR~\cite{suetal2023one} for embedding purposes, an instruction-fine-tuned encoder designed to produce task-aware embeddings. INSTRUCTOR conditions the embedding process on natural-language instructions that specify the intended task, enabling it to bring semantically aligned inputs closer while pushing apart unrelated ones. This design makes it particularly well-suited for IR and relevance assessment when paired with carefully chosen task instructions. To our knowledge, this is the first use of INSTRUCTOR embeddings for analyzing agreement between human and LLM relevance judgments.  

Since INSTRUCTOR allows the embedding to adapt to different instructions, we systematically explored nineteen alternative phrasings that covered binary classification, judgment-oriented tasks, semantic similarity, and retrieval-oriented themes. Each instruction variant produced a set of embeddings that we clustered with HDBSCAN~\cite{campello2013density}. Importantly, clustering was applied to the semantic representations rather than the relevance labels. The binarized human labels were attached only after clustering to assess how judgments distribute within each neighborhood. To evaluate embedding quality, we measured cluster purity, distinguishing clusters dominated by relevant versus non-relevant labels and inspecting their purity histograms. For cross-instruction comparison, we used the 80th percentile of cluster purity as a robust criterion, representing the purity level that at least 80\% of clusters achieve. For example, with the instruction \textit{"Judge the document’s relevance to the query for ad-hoc retrieval"} and INSTRUCTOR-XL, the 80th-percentile purity reached $\approx0.70$ for non-relevant clusters and $\approx0.61$ for relevant clusters. Since this instruction yielded stronger and more stable alignment with human judgments than alternatives (for both INSTRUCTOR-Large and INSTRUCTOR-XL), we selected it together with INSTRUCTOR-XL as the embedding configuration for all downstream analyses.

To validate this choice, we also tested alternative models. E5-Large-v2~\cite{wang2022text}, while effective in asymmetric retrieval, was unsuitable as our analysis requires joint Q–D embeddings. Qwen3-Embedding-8B~\cite{zhang2025qwen3}, despite strong benchmark performance, produced less consistent clusters with weaker majority-label alignment. These results confirmed that INSTRUCTOR-XL with the selected instruction as the most stable and effective configuration for our study.

\subsection{Clustering}
Following the embedding step, we apply HDBSCAN to cluster Q–D pairs, as it can identify dense regions in the embedding space without assuming spherical or uniform shapes. This makes it well suited to semantically diverse and unevenly distributed Q–D embeddings. Importantly, HDBSCAN assigns some points to a noise cluster ($C_{-1}$), capturing Q–D pairs that are semantically ambiguous, atypical, or poorly connected to any dense neighborhood. Rather than being discarded, such cases provide useful signals for detecting bias-prone or unstable regions where human and LLM judgments are most likely to diverge. 

\begin{figure}
  \centering
  \includegraphics[width=0.65\linewidth]{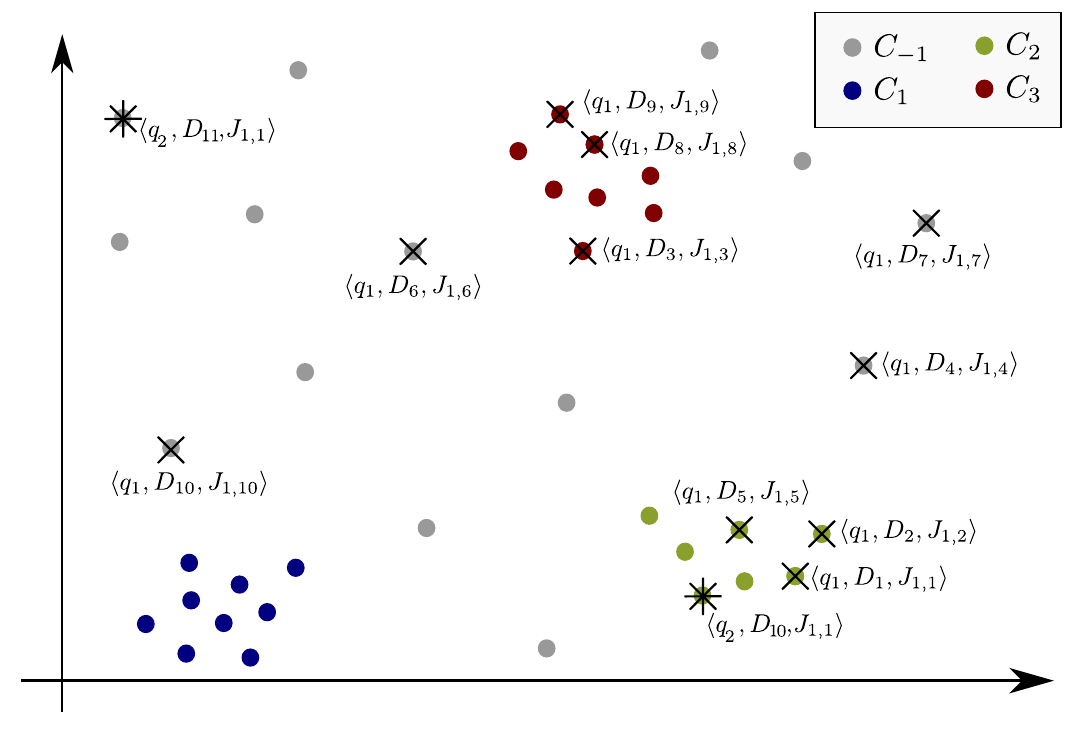}
  \vspace{-1em} 
  \caption{Clustering of query–document (Q–D) embeddings with HDBSCAN. Colored groups ($C_{1}$, $C_{2}$, $C_{3}$) denote dense semantic neighborhoods, while grey points ($C_{-1}$) represent noise. Crosses and stars respectively highlight Q–D pairs for query $q_{1}$ and $q_{2}$, illustrating how documents and queries can fall into different clusters depending on their semantic alignment.}
  \label{fig:qd_embedding_clustering}
\end{figure}

Figure~\ref{fig:qd_embedding_clustering} illustrates an example clustering of Q–D pairs. Each colored group ($C_{1}$, $C_{2}$, $C_{3}$) represents a dense semantic neighborhood, while gray points ($C_{-1}$) denote noise assignments. Crosses mark example Q–D pairs for query $q_{1}$, which appear in different clusters depending on the semantic relation of their associated documents. For instance, $\langle q_{1}, D_{1} \rangle$ and $\langle q_{1}, D_{3} \rangle$ fall into separate clusters, reflecting their distinct semantic alignments. This demonstrates how clustering separates coherent subsets of a query’s documents rather than grouping all documents of the same query together. Notably, the same document can appear in different clusters with different queries (e.g., $q_{2}$, star markers), indicating that Q–D embeddings capture semantic context rather than being dominated by the document content. After clustering, we attach binarized human and LLM relevance labels to cluster members, which enable us to assess both global and local agreement patterns. These assigned judgments are illustrated as $J_{q,d}$ like $\langle q_{1}, D_{1}, J_{1,1} \rangle$ in Figure~\ref{fig:qd_embedding_clustering}.
\subsection{Agreement Metrics}
We measure alignment between human and LLM judgments using Gwet’s AC1 as our primary agreement metric. While $\kappa$ remains the dominant measure in IR, it collapses under class imbalance, which is typical in relevance assessment. AC1 instead sets its baseline on disagreement rather than chance agreement, providing more stable estimates when labels are skewed. Evidence from medical and clinical studies shows that AC1 performs consistently under imbalance~\cite{wongpakaran2013comparison}, and simulation work further confirms its robustness across a range of binary scenarios~\cite{silveira2023better,minozzi2022kappa}. These properties make AC1 well suited for IR evaluation, where relevant documents are sparse and our setting where semantic clusters can be highly imbalanced.

\begin{figure}
    \centering
    \includegraphics[width=0.6\linewidth]{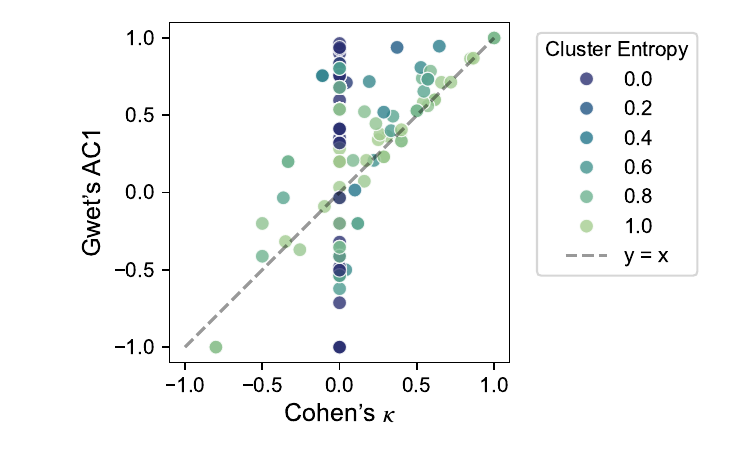} \vspace{-1.0em}
    \caption{Agreement scores of Cohen’s $\kappa$ and Gwet’s AC1 for GPT-4o judgments on DL-2019. Each point corresponds to a cluster from the HDBSCAN results.}
    \label{fig:kappa_vs_gwet}
\end{figure}

To illustrate this effect, Figure~\ref{fig:kappa_vs_gwet} shows agreement scores for an example clustering on DL-2019 with GPT-4o relevance judgments. Each point corresponds to a cluster, with color indicating label entropy as a measure of balance. Darker points represent clusters with low entropy and thus high imbalance. In these regions, Cohen’s $\kappa$ collapses toward zero while Gwet’s AC1 does not collapse under imbalance and continues to produce meaningful variation across these darker points. AC1 captures some level of agreement rather than defaulting to zero. This highlights both the prevalence of imbalance in cluster-level analysis and the instability of $\kappa$, reinforcing our choice of AC1 as the primary metric.
\vspace{-1\baselineskip}
\subsection{Cluster-Based Agreement Variation}\label{sec:agreement_variation}
Building on the clustering and agreement setup, we define the measure of \textit{Cluster-based Agreement Variation}. For query $q$, we consider all clusters that contain its associated Q–D pairs and compute Gwet’s AC1 between human and LLM judge $j$ within each cluster. This produces a set of per-cluster agreement values $\{ AC1_{j}(q,c) \}$, where $c$ indexes the clusters containing $q$. We then define the variation as the range of these values, i.e.,  
$\Delta AC1_{j}(q) = \max_{c} AC1_{j}(q,c) - \min_{c} AC1_{j}(q,c)$.  
This metric captures how strongly agreement for a given query varies across different semantic neighborhoods, highlighting queries where judgments are stable in some clusters but diverge sharply in others.

A single query $q$ can be distributed across multiple clusters depending on semantic relations, as illustrated in Figure~\ref{fig:qd_embedding_clustering}. For instance, documents $d_{1}$ and $d_{3}$ associated with $q_{1}$ fall into different clusters, so agreement between human and LLM judgments must be evaluated separately within each cluster; a large $\Delta AC1_{j}(q)$ then signals divergent alignment across contexts and exposes semantic regions that remain hidden under global agreement measures.
\vspace{-1\baselineskip}
\subsection{Heuristic Bias Localization}
Building on the variation measure $\Delta AC1_{j}(q)$ introduced in Section~\ref{sec:agreement_variation}, we define a heuristic to flag bias-prone query–judge pairs. The goal is to turn variation into a diagnostic tool that identifies items driving systematic disagreement. For each query $q$ and LLM judge $j$ in $(q,j)$, the variation $\Delta AC1_{j}(q)$ is compared against a hybrid condition, and a Q–D pair is flagged as bias-prone if it satisfies any of the following: 

1) $\Delta AC1_{j}(q) \ge \tau_{\mathrm{abs}}$, an absolute cutoff that captures variations large enough to indicate bias regardless of dataset-specific variation;  
2) $\Delta AC1_{j}(q) \ge \operatorname{median}(\Delta) + 1.5\,\operatorname{IQR}(\Delta)$, a robust data-adaptive cutoff where $\Delta$ is the set of all variations, $\operatorname{median}(\Delta)$ their median, and $\operatorname{IQR}(\Delta)$ their interquartile range. This flags queries that are statistical outliers relative to the global variation distribution;  
3) $\max_{c} AC1_{j}(q,c) > 0.8$ and $\min_{c} AC1_{j}(q,c) < 0.2$, which capture directional flips where agreement shifts from very high in some clusters to very low in others, signaling extreme instability.  

We then aggregate these local flags at the query level, marking a query as bias-prone if it is flagged by multiple judges (two or more) or by a majority of judges (more than 50\%).  To rank flagged queries, we define the \textit{Bias Severity Score} as 
$\text{BSS}_{j}(q) = \mathbbm{1}_{\mathrm{D}}(q) + \Delta AC1_{j}(q)$, 
where $\mathbbm{1}_{\mathrm{D}}(q)$ is the indicator function that equals $1$ if query $q$ is diagnosed with directional bias, and $0$ otherwise.

\section{Experiments and Results}
This section presents our experiments addressing~\ref{rq:embedding} and~\ref{rq:bias}. We first examine global agreement patterns, then localize disagreement at the query and cluster levels, and finally diagnose bias-prone queries and analyze their patterns. 
This structure reflects our contributions (\ref{contrib:c1}–\ref{contrib:c3}), moving from global statistics to fine-grained, context-aware diagnostics.

\subsection{Global Bias Analysis}
We begin with a \textit{global} perspective to assess whether LLM judges exhibit consistent bias relative to humans and how stable that bias is across the dataset. For this, we use Bland–Altman (BA) plots~\cite{altman1983measurement}, a standard method for comparing two measurement procedures. BA captures systematic bias as the average difference between methods, showing if one consistently scores higher or lower, and it captures variability through the limits of agreement(LoA), which indicate how much individual differences fluctuate. In our setting, BA reveals whether LLM–human agreement differs systematically from human–human agreement and whether such differences are stable across queries.

\begin{figure}
  \centering
  \includegraphics[width=0.88\linewidth]{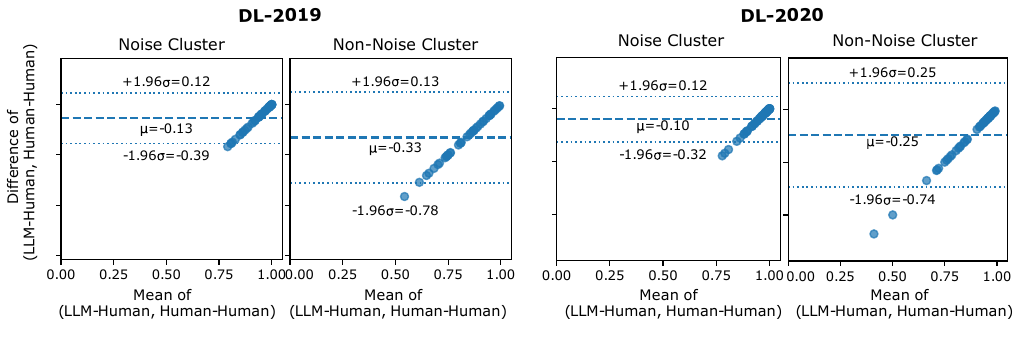}
 \caption{Bland–Altman plots for GPT-4o on DL-2019 and DL-2020 under Noise and Non-Noise cluster conditions. Each point represents a query, with the dashed line indicating mean bias and dotted lines showing the limits of agreement (LoA).}
  \label{fig:BA_plots}
\end{figure}

We analyze two conditions by partitioning the clusters for each query, if applicable, into the noise cluster ($C_{-1}$) and the non-noise clusters. Query-level AC1 is computed for each, and when queries span multiple non-noise clusters, scores are pooled by weighted averaging based on cluster size. The aggregated query-level scores are then used to derive BA estimates of bias and LoA, as shown in Figure~\ref{fig:BA_plots}.

Results reveal a consistent contrast. In the noise-cluster condition, bias is smaller and LoA tighter, indicating that sparse semantic regions are easier and more consistently judged. In the non-noise condition, bias is larger and LoA wider, suggesting that dense semantic neighborhoods drive systematic divergence. Figure~\ref{fig:BA_plots} illustrates this pattern for GPT-4o on DL-2019 and DL-2020. These findings indicate that global bias concentrates in semantically rich regions of the embedding space, motivating a shift from global summaries to query-centric diagnostics to localize sources of disagreement more precisely.
\vspace{-1em}
\subsection{Query-Level Bias Localization}
\begin{figure}[!]
  \centering
  \includegraphics[width=1.0\linewidth]{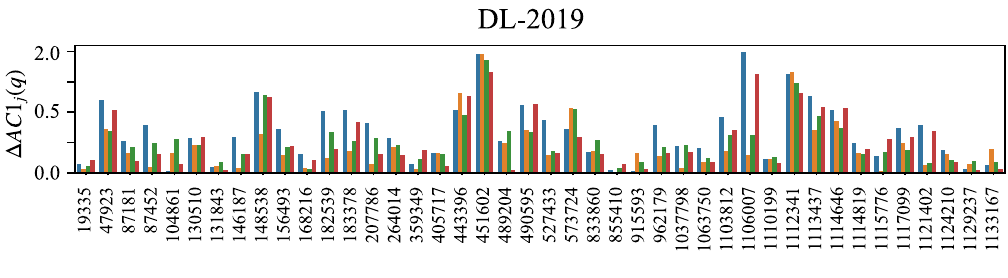}\\
  \includegraphics[width=1.0\linewidth]{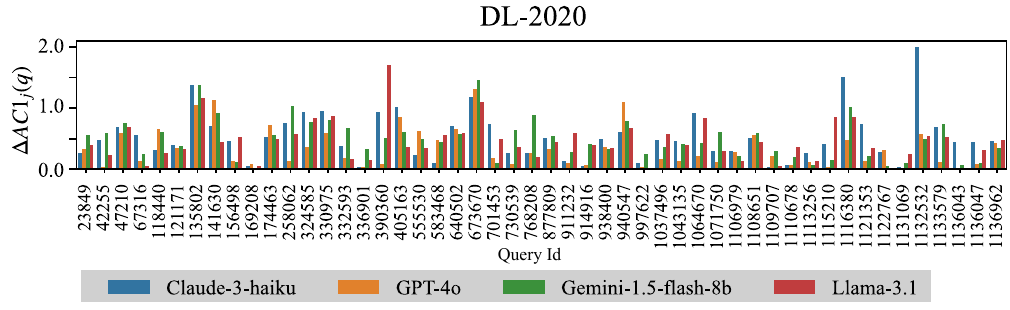}
  \vspace{-1.2em} 
  \caption{Cluster-Based Agreement Variation $\Delta AC1_{j}(q)$ calculated for all queries}
  \label{fig:gwet_variation_range}
\end{figure}

Moving beyond global summaries, we adopt a query-level perspective to identify which queries drive systematic disagreement. We use the cluster-based agreement variation $\Delta AC1(q)$ introduced in Section~\ref{sec:agreement_variation} to assess how human–LLM agreement shifts across the semantic neighborhoods of individual queries. Figure~\ref{fig:gwet_variation_range} presents the distribution of $\Delta AC1(q)$ values for different judges on the DL-2019 and DL-2020 datasets. High values indicate queries where agreement varies sharply across clusters. Such cases reveal context-dependent weaknesses in LLM assessments while low values correspond to queries with stable agreement across clusters, which suggests more robust behavior. These findings show how cluster-aware analysis provides a finer-grained view than global agreement measures and highlight queries most prone to semantic bias across both datasets.
\subsection{Heuristic Diagnosis of Bias-Prone Queries}\label{sec:bias_prone_queries}
\begin{figure}[htb]
  \centering
  \includegraphics[width=1\linewidth]{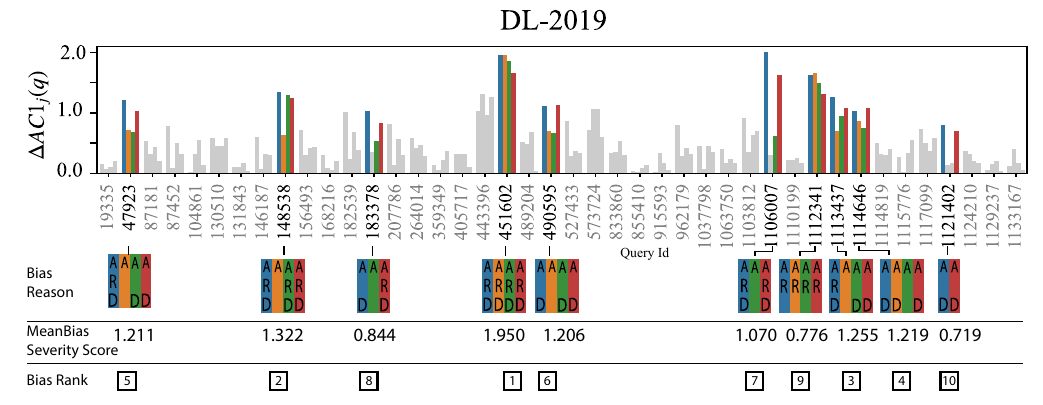}\\
  \includegraphics[width=1\linewidth]{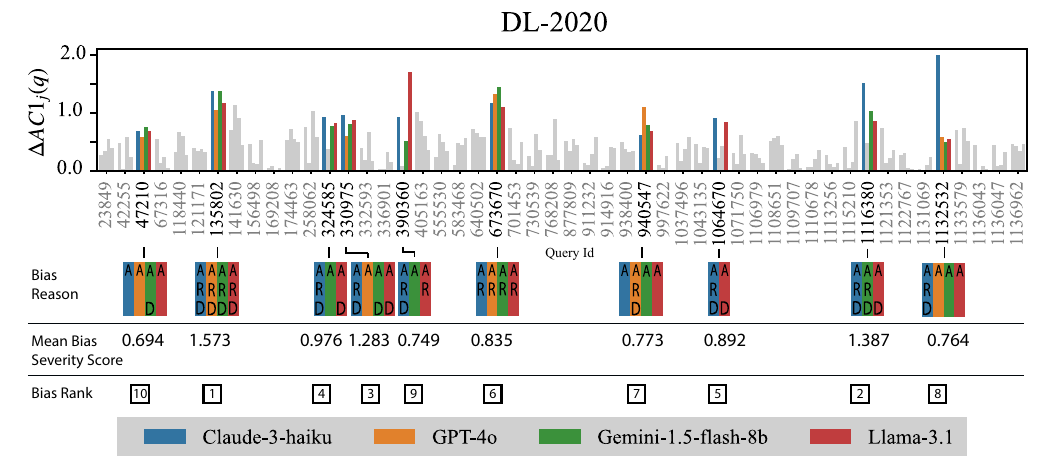}
  \vspace{-1.7em}  \caption{Top-10 bias-prone queries in DL-2019 and DL-2020 identified by the heuristic }
  \label{fig:heuristic_bias_results}
\end{figure}

We apply the heuristic defined in Section~\ref{sec:agreement_variation} to cluster-based agreement variations. A query is flagged as bias-prone if it is identified by at least two judges. We use a conservative absolute cutoff of $\tau_{\mathrm{abs}}=0.5$. Since AC1 ranges from –1 to 1, most practical values lie between 0 and 1. A deviation of 0.5 covers half of this effective range, representing a large and systematic shift. This ensures that only queries with substantial instability are flagged, prioritizing precision (capturing truly unstable cases) over recall (including minor fluctuations).
Figure~\ref{fig:heuristic_bias_results} shows the top 10 bias-prone queries in DL-2019 and DL-2020 identified by the heuristic, ranked by their mean Bias Severity Score across judges. Each bar represents the variation $\Delta AC1_{j}(q)$ for a judge. The panels below summarize the dominant bias reason, where labels A (absolute threshold), R (robust outlier), and D (directional bias) indicate the heuristic condition under which the query was flagged. Since the absolute threshold applies globally, all top queries receive an (A) flag, what differentiates them is whether they also exhibit robust outlier behavior (R) or directional flips (D).

The Bias Severity Score determines the bias rank by combining directional bias with the magnitude of variation. Highly ranked queries often accumulate all heuristic flags, reflecting strong and systematic instability. This diagnosis provides a concrete shortlist of bias-prone queries that drive model–human disagreement and forms the basis for uncovering broader bias patterns. 

This range-based heuristic provides a practical tool for identifying actionable cases of bias. It complements the Bland–Altman analyses by moving from descriptive summaries to a concrete mechanism for detection. It isolates the queries and model judgments most likely to distort evaluation outcomes and provides a basis for qualitative analysis of the semantic factors that drive disagreement.
\subsection{Qualitative Analysis of Bias Patterns}\label{sec:bias_patterns}
Building on the results of Section~\ref{sec:bias_prone_queries}, we analyze the top-10 bias-prone queries per dataset to examine the sources of disagreement and the structured patterns of bias within semantic clusters. Across both DL-2019 and DL-2020, we find that LLM failures are not random but follow recurring patterns, most often involving definitional rigidity, contextual ambiguity, and policy or role mismatches.

In DL-2019, definitional failures were the most common. Definition-seeking, informational queries (e.g., “physical description of spruce,” “famvir prescribed for”, “axon terminals”, “RSA definition key”, “exons definition biology”) showed models enforcing overly strict dictionary-style interpretations, rejecting descriptive or applied passages that humans marked relevant, while sometimes including incorrect lexical matches such as ‘terminal’ in finance. Policy and coverage mismatches arose when queries required broader policy definitions but models defaulted to narrow technical ones. For example, in “Medicare’s definition of mechanical ventilation”, models judged CPAP and BiPAP coverage passages as non-relevant while humans considered them relevant under Medicare’s framing, producing strong directional flips. Professional role distinctions and domain ambiguity also proved problematic. In “RN vs BSN”, models collapsed subtle differences that humans judged relevant, while in queries such as “daily life of Thai people” and “contour plowing”, humans flexibly accepted multiple framings but LLMs applied brittle heuristics, leading to unstable judgments across clusters.

In DL-2020, failures again followed recurring patterns. Definitional queries such as “definition of admirable”, “ia suffix meaning”, and “what is a alm” showed LLMs applying overly narrow dictionary or technical interpretations. They under-recalled broader passages that humans judged relevant and sometimes drifted into unrelated domains, for example misreading “alm” as “alms”. Domain ambiguity and cultural framing also caused sharp instability. In “nonconformity in earth science”, humans consistently marked geology passages as relevant, but LLMs misclassified them as social opposition rather than geological unconformity. In queries such as “when did rock n roll begin?” and “why do hunters pattern their shotguns?”, humans accepted multiple overlapping framings, while LLMs forced single fixed answers or admitted loosely related content. This produced cluster-level flips. Applied information needs such as “wind turbine installation cost”, “wedding dress alteration cost”, and “motivational speaker salary” showed another consistent weakness. Humans treated descriptive or context-rich passages as relevant, while LLMs either restricted judgments to narrow numeric mentions or included unrelated financial material. These failures shared the same root causes seen in DL-2019. Over-strict filtering led to systematic under-recall, and over-general lexical matching caused over-inclusion of irrelevant content.

Across both datasets, two root causes appear consistently. Over-strict filtering suppresses recall of valid contexts, while over-general lexical matching admits irrelevant material. Claude-3 and LLaMA-3.1 showed the largest divergences, often with near-zero or negative AC1 in definitional and ambiguous queries. GPT-4o and Gemini-1.5 were more stable but consistently leaned toward under-recall. 
\section{Conclusions and Implications}
This work examined how joint query–document embeddings and clustering can reveal systematic bias in LLM-based relevance judgments. Across DL-2019 and DL-2020, we found that disagreement is not random but follows structured patterns concentrated in semantic clusters. Among the top-10 analyzed bias-prone queries in each dataset, definitional rigidity, contextual ambiguity, and policy or role mismatches produced recurring errors such as under-recall, over-strict filtering, and cluster-level flips.

These observations have clear implications for the use of LLMs in IR evaluation. LLMs are well-suited for generating scalable judgments in stable regions of the embedding space, where alignment with human assessments is strong. They are less reliable for definition-seeking, policy-related, or ambiguous queries, where instability dominates and evaluation outcomes risk distortion. While our clustering-based framework requires human judgments for validation, the same signals that reveal bias, such as cross-domain overlap, definitional framing, or semantic ambiguity, can in principle be used as a pre-screening tool without the need of any human labels. This would allow practitioners to anticipate which queries are likely to induce disagreement, reducing annotation costs by applying human assessment only to unstable cases while relying on LLMs for the stable majority. Future work can extend this framework to additional IR tasks, datasets, and model families, enabling more bias-aware evaluation practices and more reliable integration of LLM-based judgments into test collection construction.

\bibliographystyle{splncs04}
\bibliography{references/references} 

\end{document}